# Proactive SRV Spectrum Handoff protocol based on GCS scheme in Cognitive Radio Adhoc Network

Morteza Mehrnoush, V. T. Vakili
*Department of Electrical Engineering,*
*Iran University of Science and Technology, Narmak, Tehran, IRAN*
mortezamehrnoush@elec.iust.ac.ir
Vakily@iust.ac.ir

*Abstract*—Cognitive radio technology allows the secondary users to utilize the spectrum, when it is not occupied by the primary users. Whenever a primary user wants to utilize a channel which is occupied by a secondary user, the secondary user should perform a proactive spectrum handoff to another channel and vacate the selected channel before the primary user utilizes it. This scheme avoids collision between the primary users and secondary users; moreover, increases the throughput of the primary and secondary users. In this paper, a novel proactive spectrum hand protocol based on the Greedy Channel Selection (GCS) is proposed which avoids collision between secondary users, as well as, collision between primary and secondary users. In the proposed scheme, proactive spectrum handoff is based on the SRV (Single Rendezvous) coordination scheme; therefore, the secondary users perform proactive spectrum handoff without using the common control channel. Moreover, the channel selection is distributed which leads to the higher throughput, and lower average service time. The proposed proactive spectrum handoff protocol is compared with the other proactive spectrum handoff protocols. Simulation results illustrate that the proposed protocol outperforms the other protocols regarding higher average throughput and lower average service time.

## I. INTRODUCTION

By increasing demand for the spectral resources, cognitive radio network emerges as a way to improve the overall spectrum usage by exploiting the spectrum opportunity [1]. Cognitive radio network allows the secondary users to use the channel whenever the channels do not occupy by the primary users. Successful deployment of cognitive radio network requires secondary users to guarantee minimal interference to primary users [2]. There are four important functionalities in cognitive radio networks: spectrum sensing, spectrum management, spectrum sharing, and spectrum mobility or spectrum handoff [3]. This paper focused on spectrum handoff in cognitive radio network. Spectrum handoff in cognitive radio network arises because primary users appear in the channel which is utilized by secondary users. Spectrum handoff allows a secondary user to vacate its current channel, when a primary user wants to initiate a new transmission in this channel, and access to a new channel for resuming the unfinished transmission [4].

There are two kinds of spectrum handoff mechanisms in the cognitive radio networks. First kind of spectrum handoff is called reactive spectrum handoff [5]. In this scheme, the target channel is searched based on demand; therefore, whenever a spectrum handoff is requested by a secondary user, spectrum sensing is started to find an idle channel for secondary user to resume its unfinished transmission. Since there is a sensing and reconfiguration delay, this scheme constrained extra delay to the network; moreover, results in collisions to both primary and secondary user transmission. Second kind of spectrum handoff is proactive spectrum handoff [6]. In this scheme, secondary users predict the appearance of primary users in the current channel of secondary users and make decision for performing a proactive spectrum handoff. Then, secondary user switch to a new channel before primary user occupy the channel. Therefore, in this scheme the collision between secondary and primary user decreases. This scheme uses the past channel usage history information to predict the future channel usage information [6, 7].

Comparison of the reactive and proactive spectrum handoff has been presented in [8]. In [7], proactive spectrum handoff scheme has been presented which the secondary users utilize greedy channel selection (GCS) method. In this scheme the channel is selected based on channel usage information and prediction of service time on each channel. This scheme considers only one pair of secondary users in the network which causes inordinate collisions between secondary users in multi user network [9]. In [10], a proactive spectrum handoff protocol based on time estimation is proposed which reduces the communication disruption and improves the channel usage, but only one pair of secondary users is considered in the network. In [9], a proactive spectrum handoff protocol based on probability prediction is proposed which by using the rendezvous coordination scheme eliminate the use of common control channel.

In this paper, a proactive spectrum handoff protocol for multi user cognitive radio ad hoc networks is proposed which selects the channel for performing spectrum handoff based on greedy channel selection. This scheme increases the average throughput of the secondary users, because in this scheme the collision between the secondary users and primary users decreases. Furthermore, the channel selection causes minimum service time for packet transmission. The secondary users which want to perform spectrum handoff or start communication should coordinate with each other for accessing the channel. In this way, the collision between the secondary users in multi user networks is avoided [11]. Since CCCs (Common Control Channels) [12] face several problems such as control channel saturation, robustness to PU activity, CCC coverage range (scalability), and control channel security in cognitive radio network, various schemes has been proposed for network coordination in cognitive radio networks without CCC [13]. In the proposed scheme the secondary users coordinate with each other to perform proactive spectrum handoff in cognitive radio ad hoc network by using the single rendezvous scheme.

## II. Proposed Spectrum Handoff Model

In this part, initially network assumption is stated. Then, channel selection scheme for selecting the best channel for performing spectrum handoff is presented. Moreover, the criteria for selecting the best channel are introduced in this part.

### A. Assumption

In this paper, the model of channel is assumed as an ON-OFF process. The hachure rectangles show the primary users packet transmission as an ON process and other area are the OFF process which show the primary users do not have any packet for transmission. The primary and secondary users are M/G/1 systems and the packet arrival rate of both secondary and primary users follow the poisson distribution process [6, 8]. Average arrival rate of the primary users and secondary users are λp and λs, respectively.

By considering that the power of the transmitted signal is higher than the received signal, the instantaneous collision detection and transmission is not possible for wireless nodes. Therefore, in the proposed scheme, it is assumed that each secondary user is provided with the two radios [14, 15]. First kind of radio is called the transmitting radios and is used to transmit the data and control packets. Second kind of radio is called the scanning radio and is used to scan the channel and gather the channel usage information. Moreover, the scanning radio should have the ability to scan the selected channel to certify the selected channel is not occupied by another secondary user.

### B. Channel Selection Scheme

In this section, the proposed scheme for proactive spectrum handoff is presented which is based on multi-user improved greedy channel selection. In this scheme, all secondary users predict the service time on each channel based on channel usage information. Secondary users for initiating a spectrum handoff should select the best channel based on the two bellow criteria. First, selecting the best channel based on minimum service time. Second, selecting the best channel based on maximum vacant time. Fig. 1, shows a sample model of improved greedy channel selection scheme in the multi-user cognitive radio network which secondary users select the best channel based on two proposed criteria. There are M channel and K secondary users in the network.

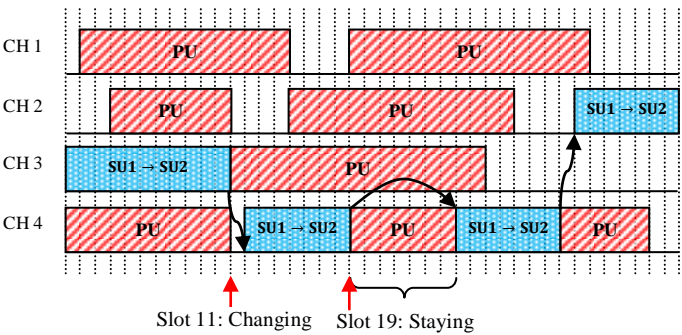

Fig. 1. Both secondary and primary user transmits packet. Secondary user performs spectrum handoff when primary user wants to occupy the channel.

In Fig. 1, the SU-1 starts the transmission to SU-2 in channel 3 in the first slot. After 11 slots the PU-3 appears in channel 3 and wants to initiate a new transmission. Hence, the secondary user should decide to change the channel and perform a spectrum handoff, or stay in the current channel and resume its unfinished transmission after the primary user vacates the channel. In this sample model, the secondary user changes its channel and resumes the transmission on channel 4, because this channel has minimum service time and maximum vacant time. In slot 19, the primary user also appears in channel 4 and the secondary user should decide to stay in the current channel, or change its channel and performs a spectrum handoff. Thus, the secondary user decides to stay in its current channel and continue the unfinished transmission after primary user finished its transmission.

Secondary user for selecting the channel compares the stay time in its channel with the changing time of other channels:

$$ST_{i,j} = \begin{cases} S_k & if\ S_k \leq C_j + t_h \\ C_j & if\ S_k \geq C_j + t_h \end{cases} \quad (1)$$
$$for\ j = 1, \dots, M$$

Where $S_k$ denotes the staying time in the current channel of the secondary user ($k$ is the number of channel which user $i$ utilizes it), and $C_j$ is the changing time of the channel which secondary user could select. Let $t_h$ denotes the handoff delay for changing its channel to another channel ($j$). If the staying time is less than the changing time the secondary user stays in its channel. But if the changing time plus handoff delay is shorter than the staying time the secondary user decides to change its channel and use the second criterion for best channel.

As the second criterion, the secondary user should select the channel with maximum vacant time among all channels which have zero changing time:

$$VT_m = \underset{m \neq k}{sort}(T_m)\ for\ m = 1, \dots, M\ \&\ m \neq k \quad (2)$$

where $T_m$ is the vacant time of the channels which has the zero changing time. It is the time slot from the instant the secondary user wants to perform a spectrum handoff until primary user wants to occupy the channel. By considering these two criteria, secondary users can select the best channel.

## III. SRV Spectrum Handoff Protocol

In this section, split phase coordination protocol in cognitive radio network is explained. Moreover, proactive spectrum handoff scheme in SRV coordination protocol when a primary user wants to occupy the secondary users channel is illustrated.

### A. Split Phase Coordination Protocol

In this paper, split phase coordination protocol is used as a single rendezvous coordination scheme [16, 17]. In this approach, time is divided in to an alternating sequence of data and control phases. During control phase all devices tune to the control channel and try to make agreement to access the channels in following data exchange phase [17, 18]. In the transmitted phase, secondary users tune to the agreed channel and start data transfer. Fig. 2, shows the operation of the split phase coordination scheme in the cognitive radio network.

According to considered scenario, there are M orthogonal channels in the network and K secondary users content to access the channel. For coordinating between the secondary users, each time slot is segmented to 2K mini slot. The secondary user for accessing the channel should contend with other secondary users in control phase. When a secondary user wants to start a new transmission or perform a spectrum handoff, initially, it sends a RTS (Ready-To-Send) packet to the intended receiver in control phase. If the receiver agrees with the selected channel by the transmitter, replies with CTS (Clear-To-Send) packet. Then, they hop to the selected channel and start their transmission. If the secondary users cannot make agreement for selecting the channel, they should wait until the next control phase and send RTS/CTS in the next control phase for accessing the channel.

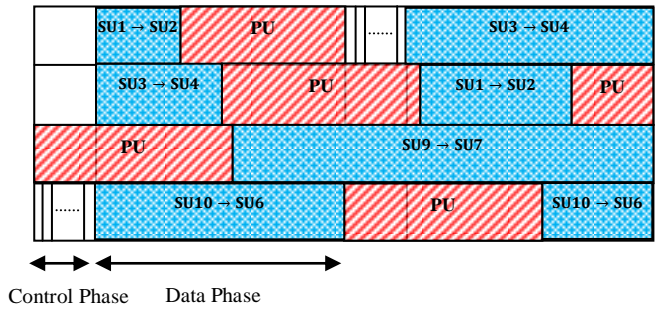

Fig. 2. Split Phase coordination protocol as a SRV Proactive Spectrum Handoff Protocol.

In Split Phase coordination protocol the secondary users need to be synchronized. This scheme needs time synchronization between all of the secondary users. Time synchronization in this protocol can be looser, since in this scheme secondary users hops fewer than the other schemes such as common happing and MMAC. Time synchronization scheme is used in this paper is similar to the scheme is presented in [19].

### B. SRV Proactive Spectrum Handoff Protocol

In SRV coordination scheme, when a secondary user wants to start transmission or to carry out a spectrum handoff, it should predict and select the channel based on two proposed criteria. The secondary users should tune to the best channel in each control phase as a rendezvous channel for coordination. Rendezvous channel has the minimum delay and maximum vacant time. In control phase, secondary users which do not want to carry out a spectrum handoff can continue their transmissions. Secondary users before sending control packets and at the start of the control phase should scan the rendezvous channel, because some of the channels are occupied by other secondary users from previous data phase. If the Rendezvous channel is occupied by a secondary user, the users which want to coordinate should go to next best channel for rendezvous at the same time slot and scan it. When the Rendezvous channel is not occupied by a secondary user, secondary users can send their control information packets.

The secondary users which want to start packet transmission should tune to the rendezvous channel and sends a RTS packet on the control phase. RTS packet contains the number of selected channel by secondary user transmitter. Secondary user receiver upon receiving RTS packet, if agrees with the selected channel, replies with a CTS packet at the same mini time slot. If the secondary user transmitter receives the CTS packet, two secondary users tune to the selected channel and start data transmission.

The secondary user for performing a proactive spectrum should use the proposed channel selection criteria and coordinate with other secondary users in the control phase. For coordinating with other secondary users in the control phase, the secondary user transmitter sends a RTS packet to the receiver which it contains the newly selected channel. If the secondary user receiver is agree with selected channel replies with the CTS packet. After receiving the CTS packet by secondary user transmitter, they switch to the selected channel and continue the transmission. If there is no channel that the secondary user can select for transmission, it should wait until next control phase to select the channel. Pseudo-code[1] of the GCS proactive spectrum handoff SRV protocol is presented in this section.

---

**GCS Proactive Spectrum Handoff SRV Protocol**
Initially: LSC=∅, LNC=∅, FDT=1, FCS=0, FDS=0
For j=1, j≤M
    Determine $ST_{i,j}$ and $VT_m$;
    If $ST_{i,j}$=0
        LNC(NT)=j;
        NT= NT+1;
    end
end
Determine the LSC based on sorting $ST_{i,j}$ and $VT_m$;
% a primary user occupy the i secondary user channel
% which is the channel k
If k = LSC(1)
    FCS(i)=0;
Elseif k ≠ LSC(1) & $ST_{i,k}$ > min($ST_{i,j}$)
    FCS(i)=1;
End
If LNC=∅
    Stop transmission until next control phase;
Elseif LNC≠ ∅
    Select the best channel for coordination;
    If channel is busy select the best next channel for coordination;
    Start scanning channel;
    Go to channel selection algorithm;
    Send RTS packet;
end
By receiving CTS packet Then
Go to the selected channel and start scanning;
If selected channel is busy
    Stop transmission until next control phase
Elseif
    FCS=0; FDS=1;
    If FDS=1
        Start data transmission;
        FDS=0;
        By finishing the data transmission: FDT=0;
    end
end

---

The secondary users that need to perform spectrum handoff in a control phase should use the distributed channel selection scheme. The channel selection should be distributed because the cognitive radio ad hoc network has a distributed entity and there is no way to manage the spectrum allocation in a centralized manner to prevent collisions. The channel selection algorithm should avoid collision among secondary users. When there is more than one pair of secondary user which wants to initiate data transmission or perform a spectrum handoff and send RTS at the same time (i.e. at the same control phase), distributed channel selection should avoid collision between them. Moreover, when there are more than one pairs of secondary users which want to carry out a spectrum at the same time, the distributed channel selection should avoid collision between them. Since collision among secondary users result in the data transmission failure, and cause a long spectrum handoff delay, it's more important in proactive spectrum handoff [15].

---

[1] LSC is the List of the Secondary users Channel, LNC is the List of the Negotiation Channels, FDT is the Data Transmission Flag, FCS is the Channel Switching Flag, and FDS is the Data Sending Flag.

In the distributed channel selection, the secondary users which want to carry out spectrum handoff should have the priority for channel selection. Because in delay sensitive application, the transmission of the secondary users should not be stop and delayed, and long spectrum handoff has a deteriorating effect. Therefore, control phase is divided in to two parts. In the first part of control, the secondary users which want to carry out a spectrum handoff can send RTS and receive CTS in their corresponding mini slot. In the second part, the secondary users that want to initiate packet transmission can send RTS/CTS for accessing the channel.

Each secondary user which wants to carry out a spectrum handoff or initiating a new transmission should generate a pseudo-random channel selecting sequence and follow it to choose the best channel. In each control phase a Pseudo-random selecting sequence is generated in all secondary users which all secondary users should follow. The selecting sequences are also different in various control phases. For selecting the best channel, the secondary users start channel selection based on pseudo-random channel selecting sequence, and the first secondary user in each control phase select the best channel based on the presented criteria. If the first secondary user selects a channel, the other secondary users should remove the selected channel from their lists. Pseudo-code[2] of the channel selection coordination algorithm is shown in the next page.

**Channel Selection Coordination Algorithm**

Input: PRS, LSC
Output: Selected Channels $C_{s(i)}$
% For secondary users that perform Spectrum Handoff
For i=1, i≤P     % Start from the first secondary user
   If FDT(PRS(i))=1
      C(PRS(i))=LCS($N_C$);
      LCS=LCS - C(PRS(i));
      $N_C$= $N_C$+1;
   end
end
% For secondary users that start transmission
For i=1, i≤P     % Start from the first secondary user
   If FDT(PRS(i))=1
      C(PRS(i))=LCS($N_C$);
      LCS=LCS - C(PRS(i));
      $N_C$= $N_C$+1;
   end
end

---

[2] PRS is the Pseudo Random channel Selecting sequence, and $C_{s(i)}$ is the selected channel for the *i* secondary user.

The secondary users should exchange the sensed channel availability information in SRV scheme. Therefore, when a secondary user wants to carry out a spectrum handoff first it broadcast the sense channel availability information to other neighboring secondary user on its corresponding mini slot in the control phase to avoid collision with other secondary users which want to broadcast the sense channel availability information.

## IV. SIMULATION RESULTS OF THE PROPOSED PROACTIVE SPECTRUM HNADOFF PROTOCOL

In this section, simulation result of the proposed proactive spectrum handoff protocol is presented. In this simulation setup, the channel bit rate is 1Mbps and each time slot takes 1ms. The data packet length of both primary users and secondary users are constant. Packet length of the secondary user is $6 \times 10^4$ and packet length of the primary user is $1 \times 10^5$.

In this simulation, it's assumed that each secondary user have the capability of prediction to predict all the channels. Performance of the proposed scheme is evaluated through simulation and compared with the other proactive spectrum handoff protocols.

Fig. 4, shows the average throughput of the proposed GCS spectrum handoff SRV protocol under the different traffic load of the primary users. It's assumed, there are 10 channels and 10 pairs of secondary users in the network and the secondary users which have packet for transmission contending with other secondary users for accessing the channels. By increasing the primary user traffic load, the opportunity of secondary users for accessing the channel decrease and therefore the average throughput of secondary users decreases.

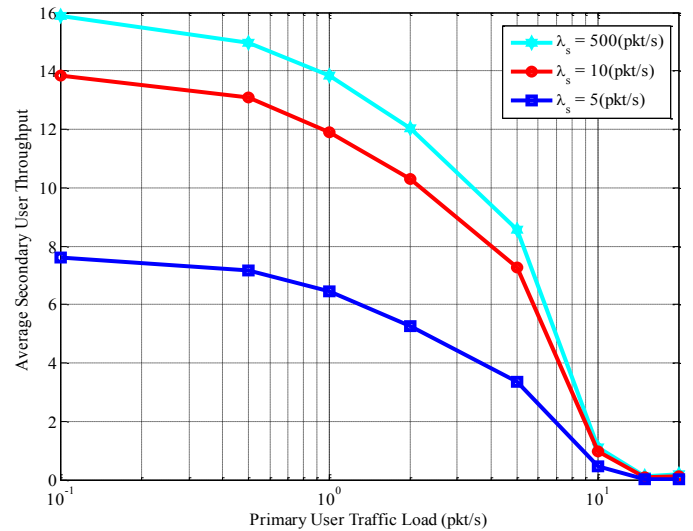

Fig. 4. Average throughput of the secondary users under different traffic load of the primary users.

Fig. 5, the proposed GCS spectrum handoff SRV protocol is compared with the random channel selection protocol and probability based spectrum handoff protocol. It's assumed that there are 10 channels and 10 secondary users in the network and the packet transmission rate of the secondary users is $\lambda_s = 500$ (pkt/s); therefore, the secondary users always have packets for transmission. As indicated in Fig. 5 the throughput of the proposed scheme is better than the two other schemes. Because in random channel selection scheme the secondary

users select their channels randomly and in probability based spectrum handoff protocol the channel selection is based on probability; hence, these schemes cause a collision between secondary users and primary users. But in proposed scheme because the secondary users predict the channels based on channel observation history there is no collision between primary users and secondary users. Therefore, average throughput of secondary users of these schemes is lower than the proposed scheme. Fig. 6, also, compares the average service time of the proposed scheme with the random channel selection protocol and probability based channel selection protocol. It can be observed that the proposed scheme causes a fewer average service time than the other schemes, since in GCS scheme the secondary users select the channels which causes minimum service time.

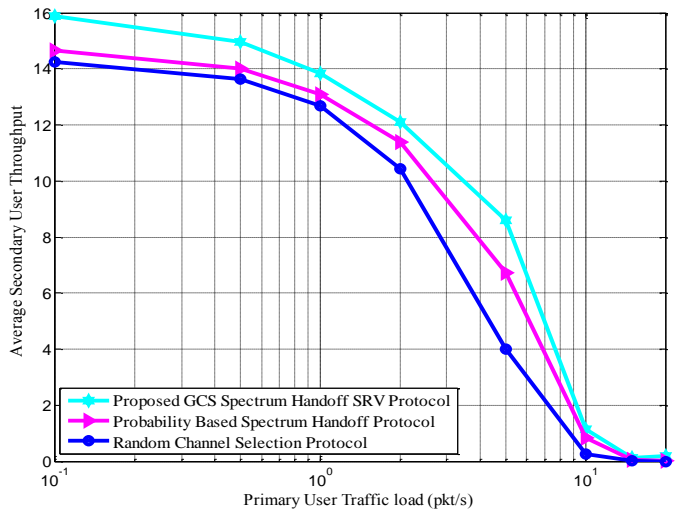

Fig. 5. Comparison of the proposed SRV greedy channel selection protocol with the random channel selection and probability based spectrum handoff protocol when there are 10 channels and 10 secondary users in the network.

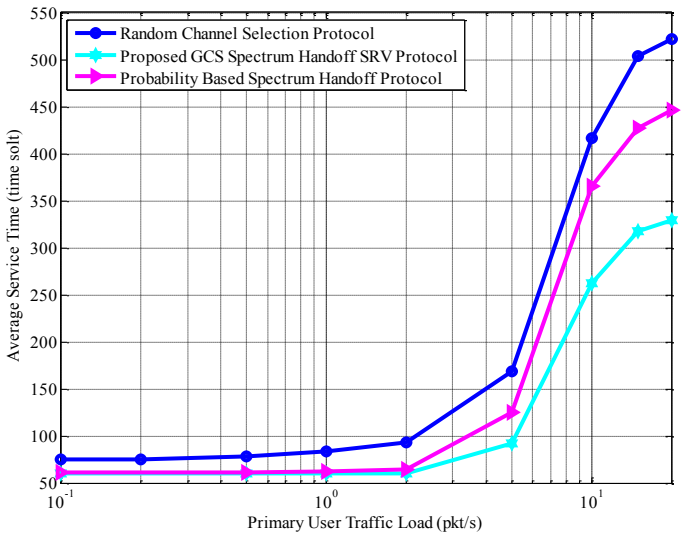

Fig. 6. Average service time comparison of proposed GCS spectrum handoff SRV protocol with the probability based spectrum handoff protocol and random channel selection protocol.

Fig. 7, shows the average secondary users throughput under different number of secondary users in the network. It's assumed that there are 20 channels in the network, the packet transmission rate of the secondary users is $\lambda_s = 500$ (pkt/s) and packet transmission rate of the primary users is $\lambda_p = 5$ (pkt/s). As the number of the secondary users increase, average throughput of the secondary users decrease, since the chance of idleness of channels for packet transmission decreases. It is, also, shows that the performance of the proposed scheme is better than the two other schemes.

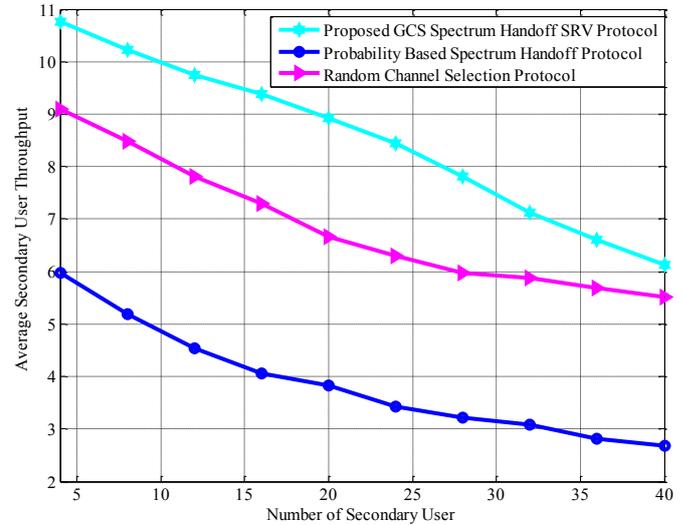

Fig. 7. Average throughput comparison of proposed GCS spectrum handoff SRV protocol with the probability based spectrum handoff protocol and random channel selection protocol under different number of secondary users.

Fig. 8, shows the average throughput of secondary users under different number of channels in the network. It's assumed that there are 10 secondary users in the network, the packet transmission rate of the secondary users is $\lambda_s = 500$ (pkt/s) and packet transmission rate of the primary users is $\lambda_p = 5$ (pkt/s). By increasing the number of channels, the average throughput of the secondary users increases. After a certain number of channels the average throughput leads to a saturate throughput, because increasing the number of channels does not provide more opportunity for the secondary users to access the channel. Average throughput of the proposed protocol is, also, better than the two other protocols.

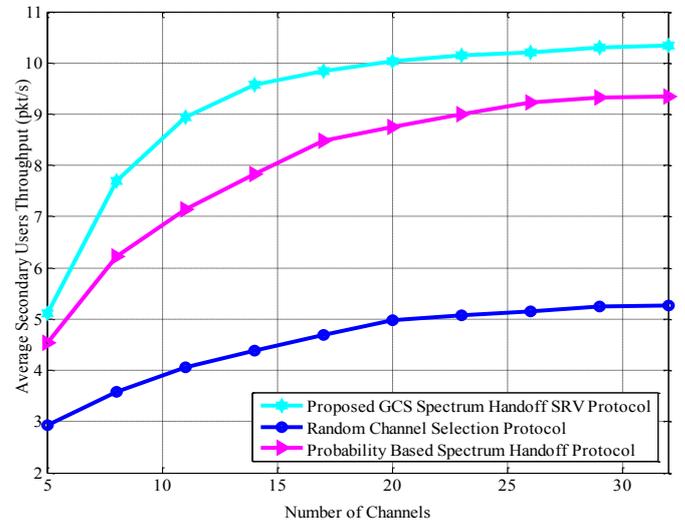

Fig. 8. Average throughput comparison of proposed GCS spectrum handoff SRV protocol with the two other protocols under different number of channels.

## V. CONCLUSION

In this paper, a novel proactive spectrum handoff protocol is proposed. The proposed protocol is based on the greedy channel selection scheme. The proposed GCS proactive spectrum handoff SRV protocol is used the split phase channel coordination protocol. Moreover, a distributed channel selection algorithm is proposed to perform channel selection in this protocol. The SRV proposed protocol is compared with the probability based spectrum handoff protocol and random channel selection protocol. The simulation result reveals that, the proposed protocol has a better performance than the other schemes from the viewpoint of average throughput and minimum service time of the secondary users.